\documentclass{ptephy_v1}

\preprintnumber{XXXX-XXXX} 

\usepackage{amsmath,amssymb}
\usepackage{bm,epsfig,ulem}
\usepackage{color}

\begin{document}

\title{Coulomb screening correction to 
the $Q$ value of the triple alpha process in thermal plasmas}


\author[$\dagger$1]{Lai Hnin Phyu}

\author[1]{H. Moriya\thanks{These authors contributed equally to this work.}}

\author[1*]{W. Horiuchi}

\affil{Department of Physics, Hokkaido University, Sapporo 060-0810, Japan \email{whoriuchi@nucl.sci.hokudai.ac.jp}}

\author[2*]{K. Iida}

\author[2]{K. Noda}

\affil{Department of Mathematics and Physics,   Kochi University, Kochi 780-8520, Japan \email{iida@kochi-u.ac.jp}}

\author[3]{M.~T. Yamashita}
\affil{Instituto de F\'isica Te\'orica,
  Universidade Estadual Paulista, UNESP, 
  Rua Dr. Bento Teobaldo Ferraz, 271 - Bloco II,
  S\~ao Paulo, SP 01140-070, Brazil} 
  
\begin{abstract}
The triple alpha reaction is a key to $^{12}$C production
and is expected to occur in weakly-coupled, thermal plasmas
as encountered in normal stars.
We investigate how Coulomb screening affects the 
structure of a system of three alpha particles in such 
a plasma environment by precise three-body calculations
within the Debye-H\"uckel approximation.
A three-alpha model that has the Coulomb interaction
modified in the Yukawa form is employed.
Precise three-body wave functions
are obtained by a superposition of correlated Gaussian bases
with the aid of the stochastic variational method.
The energy shifts of the Hoyle state due to
the Coulomb screening are obtained as a function
of the Debye screening length.   The results, which 
automatically incorporate the finite sizoe effect of the Hoyle state, 
are consistent with the conventional result based on
the Coulomb correction to the chemical potentials of 
ions that are regarded as point charges in a weakly-coupled,
thermal plasma.
We have given a theoretical basis to the conventional point-charge 
approach to the Coulomb screening problem relevant for 
nuclear reactions in normal stars by providing
the first evaluation of the Coulomb corrections 
to the $Q$ value of the triple alpha process that produces
a finite size Hoyle state.
\end{abstract}

\subjectindex{xxxx, xxx}

\maketitle

\section{Introduction}

In the past few decades, the structure of the $^{12}$C spectrum
has been one of the most interesting phenomena
in nuclear physics~\cite{Beck}.
An accurate description of the production process
of the $^{12}$C element, which is one of
the most abundant elements, is a key to understanding
the nucleosynthesis in normal stars~\cite{Ishikawa13, Suno15, Akahori15, Suno16},
where $^{12}$C is created in the fusion of three $^4$He nuclei 
($\alpha$ particles) through the formation of
the $^{8}$Be resonant state as an intermediate state~\cite{Salpeter52}.
To explain the abundance of $^{12}$C, in the 1950s,
Hoyle proposed the existence of a resonant state of $^{12}$C 
with $J^{\pi}=0^+$, the same spin-parity as the ground state,
at an energy just above the three-$\alpha$ threshold.
This state, which is called the Hoyle state~\cite{Hoyle54},
was experimentally confirmed soon afterwards~\cite{Dunbar53}
and has been believed to play an essential role
in increasing the production rate of $^{12}$C. 

From an astrophysical point of view, a dense and cold helium plasma 
also appears in the outer layer of an X-ray bursting, accreting neutron 
star~\cite{Lewin93}, in which the triple alpha reaction  
leads to unstable helium burning.
In such a plasma, the Coulomb repulsion is screened off at large 
distances by the surrounding degenerate electrons in a 
manner that is dependent on the plasma density~\cite{Salpeter54}. 
This phenomenon can affect the triple alpha reaction rate as it 
shifts the energy of the Hoyle state. 

In this paper, we study the Coulomb screening effect on
the Hoyle state in such a plasma environment as encountered 
in the normal stars that undergo a stable burning of helium.
To describe the structure of the Hoyle state of $^{12}$C,
we perform precise three-body calculations in
terms of the correlated Gaussian expansion with the aid of
the stochastic variational method~\cite{Varga95,SVM}.
In these calculations, three $\alpha$ particles are approximated 
as structureless point particles, while phenomenological two- 
and three-$\alpha$ potentials that reasonably reproduce the
empirical $^{8}$Be and Hoyle state energies~\cite{Suno15} are employed.  
Finally, the Coulomb screening effect is incorporated
into the Coulomb interaction in the Yukawa form.
This form is relevant as long as the Debye-H\"uckel 
approximation is valid.  This approximation gives a good 
description of the long-range Coulomb screening 
in a weakly-coupled, thermal plasma of interest here.

Since the screening acts to reduce the Coulomb interaction
between $\alpha$ particles, this astrophysical environment  
would make it less repulsive than that in a free space.
In fact, the screening effect in the three-$\alpha$ 
system was studied to search for the possible Efimov 
states~\cite{Jensen95, Higa08}. It was concluded that
due to the nature of the Hoyle state that appears as the 
three-$\alpha$ first excited state,
a series of the Efimov $0^+$ states might appear above 
the Hoyle state in possible astrophysical environments 
where the two-$\alpha$ ground state energy would become 
almost zero. 
According to Ref.\ \cite{Naidon17}, a full treatment 
of the three-body problem with short-range and Coulomb 
interactions could shed some light on the conjecture that 
the Hoyle state may emerge as an Efimov state.

The paper is organized as follows. In Sec.~\ref{model.sec}, 
we explain how to construct the wave function of the 
three-$\alpha$ system.  In doing so, the variational 
method and model Hamiltonian are described. 
Section~\ref{screening.sec} is devoted to a description of 
the correction to the Hoyle state energy by the Coulomb 
screening in a weakly-coupled plasma in the 
zero-size limit of ions including a $^{12}$C nucleus in the Hoyle state.  
The validity of the Yukawa form of the screened
Coulomb potential is discussed in this section.
In Sec.~\ref{results.sec}, we show the calculated results 
for the energy shift of the Hoyle state due to the
screening and compare them with those based on 
the Coulomb corrections to the chemical potential of 
point ions in weakly-coupled plasmas.
Conclusions of this work are drawn in Sec.~\ref{conclusion.sec}.

\section{Three-$\alpha$ description of the Hoyle state}
\label{model.sec}

In this section, we describe how to obtain the three-$\alpha$ 
wave function while allowing for the Coulomb screening.

\subsection{Variational calculation with correlated Gaussian expansion}

We begin by summarizing a variational approach, which will be
adopted to obtain a precise solution of the three-body Schr\"odinger 
equation.  The Hamiltonian for the three-$\alpha$ system is specified as
\begin{align} 
H = \sum_{i=1}^{3}T_i-T_{\rm cm} 
+\sum_{i<j}  [V_{ij}^{2\alpha}+V^{\rm Coul}_{ij}(C)]
+V^{3\alpha}_{123},
\label{hamiltonian}
\end{align}
where $T_i$ is the kinetic energy operator of the $i$th $\alpha$ particle,
and the center-of-mass kinetic energy $T_{\rm cm}$ is appropriately subtracted.
Details of the two- and three-$\alpha$ potentials,
$V_{ij}^{2\alpha}$ and $V_{123}^{3\alpha}$, 
as well as the screened Coulomb potential $V^{\rm Coul}_{ij}(C)$ 
with the screening factor $C$, will be given in the next section.

The wave function of the three-$\alpha$ system can be expanded by
a number ($K$) of symmetrized ($\mathcal{S}$) 
correlated Gaussian basis functions $G$ as
\begin{align}  
\Psi^{(n)}=\displaystyle\sum_{k=1}^{K} \, c_k^{(n)} \, \mathcal{S} \, G(A_k,\bm{x}).
\end{align}
The set of the coefficients $(c_1^{(n)},\dots,c_K^{(n)})$, 
where $n$ denotes a label of the state $(n=0,\dots, K-1)$
with $n=0$ being the ground state,
can be determined by solving the generalized eigenvalue equation
\begin{align}
 \sum_{j=1}^{K} H_{ij} c_{j}^{(n)} = E^{(n)}\sum_{j=1}^K \; B_{ij} c_j^{(n)},
\end{align} 
where
\begin{align}
H_{ij}&=\left<\mathcal{S}G(A_i,\bm{x})|H|\mathcal{S}G(A_j,\bm{x})\right>
\end{align}
and
\begin{align}
B_{ij}&=\left<\mathcal{S}G(A_i,\bm{x})|\mathcal{S}G(A_j,\bm{x})\right>
\end{align}
are the Hamiltonian and overlap matrix elements, respectively.

Here, the coordinate set $\tilde{\bm{x}}= (\bm{x}_1,\bm{x}_2)$,
where the tilde denotes the transpose of the matrix,
is taken as the Jacobi coordinates
excluding the center of mass of the three-$\alpha$ system $\bm{x}_3$.
These three coordinates are defined as
\begin{align}
\bm {x}_i=\sum_{j=1}^{3} U_{ij} \bm{r}_j,
\end{align}
where $\bm{r}_i$ denotes the $i$th single-$\alpha$ coordinate, and
\begin{align}
U=\begin{pmatrix}   
1    & -1   & 0    \\
\frac{1}{2}	& \frac{1}{2}	& -1   \\
 \frac{1}{3} 	& \frac{1}{3}	& \frac{1}{3}    \\
\end{pmatrix}
\end{align}
is the transformation matrix. Finally,
the correlated Gaussian basis function is defined by~\cite{Varga95,SVM}
\begin{align}
G(A,\bm{x})&=\exp\left (-\frac{1}{2}\tilde{\bm{x}}A\bm{x}\right)\notag\\
&=\exp\left(-\frac{1}{2}A_{11}x_1^2-\frac{1}{2}A_{22}x_2^2
-A_{12}\bm{x}_1\cdot\bm{x}_2\right).
\label{gaussian}
\end{align}
Each correlated Gaussian is specified by  
a symmetric, positive-definite $ 2 \times 2 $ matrix $A$.
The diagonal elements of the matrix $A$ can be related to
 the Gaussian falloff parameters as $1/\sqrt{A_{ii}}$,
while the off-diagonal element controls
the correlations among the different relative coordinates. 

The wave function of the system has to have a proper symmetry 
under interchange of identical particles.
The symbol $\mathcal{S}$ denotes a symmetrizer that assures the basis 
function being totally symmetric with respect to any exchange of particles.
One of the major advantages of the correlated Gaussian 
is invariance of its functional form under any coordinate transformation,
which allows us to easily manipulate exchange of the particles
as needed for the symmetrization of the basis function.
In fact, we superpose the six permutations among the three 
identical bosons, which in turn can be expressed by an appropriate 
choice of the transformation matrix $T_P$. 
The quadratic form $\tilde{\bm{y}}A\bm{y}$ can thus be rewritten
as $\tilde{\bm{x}}\tilde{T}_PAT_P\bm{x}$ with the
transformation of the coordinate set $\bm{y}=T_P\bm{x}$, which 
leaves the functional form of the correlated Gaussian unchanged.
This convenient property makes the correlated Gaussian basis 
suitable for treating few-body systems accompanied by
strong interparticle correlations~\cite{Mitroy13,Suzuki17}.

Most of the matrix elements, including $H_{ij}$ and $B_{ij}$, 
can be analytically obtained as functions of
a number of variational parameters, i.e., 
the matrix  elements $A_{ij}$ for each basis~\cite{Varga95,SVM,Suzuki08},
which are in turn optimized by the stochastic variational 
method~\cite{Varga95, SVM}.
In practice, the diagonal matrix elements of the matrix $A$ 
are generated as random numbers in the ranges of 
$0<1/\sqrt{A_{11}}<20$ fm and $0<2/\sqrt{3A_{22}}<20$ fm
in such a way that one can describe the asymptotic wave function
due to the Coulomb screening at large distances.
The correlation among the particles is taken into account via
the off-diagonal matrix element $A_{12}$, which is determined
by defining the two-dimensional rotation matrix $R(\theta)$ with 
randomly generated rotational angles $\theta$
and multiplying it to the diagonal matrix 
$D_{ij}=A_{ij}\delta_{i,j}$ as $\tilde{R}DR$. 

We remark in passing that the above-mentioned formalism holds also
for description of the ground state structure of $^8$Be.  In this case,
one can omit $V^{3\alpha}_{123}$ from the Hamiltonian (\ref{hamiltonian})
and set $A_{22}=A_{12}=0$ in the Gaussian basis function 
(\ref{gaussian}).

\subsection{Potential terms in the three-$\alpha$ Hamiltonian}

In describing $V_{ij}^{2\alpha}$ in Eq.\ (\ref{hamiltonian}) 
not only in as simple a form as possible but also in such a way 
as to reasonably reproduce low energy $\alpha$-$\alpha$ scattering data,
we assume the $\alpha$ particle to be 
an inert point boson.  Several versions of the potential 
models constructed under such an assumption are known 
(see, e.g., Refs.~\cite{AB, BFW, Theeten06}).
In this paper,
we employ the modified Ali-Bodmer (AB) potential~\cite{AB,Fedorov96},
which is designed to provide the $S$-wave $^{8}$Be ($0_1^+$) 
resonance position $E_r$ with 88.84 keV~\cite{Suno15}, a value close
to the empirical one 91.8 keV~\cite{Ajzenberg90}.
The explicit form of the potential is given by
\begin{align} 
V_{ij}^{2\alpha} = 125\exp\left(-\frac{r_{ij}^2}{1.53^2}\right)
-30.18\exp\left(-\frac{r_{ij}^2}{2.85^2}\right),
\label{2apot.eq}
\end{align}
where the energy and length are given in units of MeV and fm, 
respectively, and $r_{ij}\equiv|\bm{r}_i-\bm{r}_j|$.

To take the Debye screening in thermal plasmas into account,
we replace the bare Coulomb potential between point charges 
located at $\bm{r}_i$ and $\bm{r}_j$ by the Yukawa form
\begin{align}
V_{ij}^{\rm Coul}=\frac{4e^2}{r_{ij}}\exp\left(-Cr_{ij}\right).
\label{Coul}
\end{align}
Here, the parameter $C$ acts as the inverse of the length
of the Coulomb screening.  The validity of this form of 
the screened potential and the relevant value of $C$
will be given in Sec.~\ref{screening.sec}.  
We remark in passing that for more realistic calculations,
the charge form factor of an $\alpha$ particle, $f$, can be 
incorporated into Eq.\ (\ref{Coul}) as
\begin{eqnarray}
V_{ij}^{\rm Coul}&=&
\int d^3 u_i \int d^3 u_j f(\bm{u}_i)f(\bm{u}_j)
 \frac{4e^2}{|(\bm{r}_i+\bm{u}_i)-(\bm{r}_j+\bm{u}_j)|} 
\nonumber \\ & &
 \times
    \exp\left(-C|(\bm{r}_i+\bm{u}_i)-(\bm{r}_j+\bm{u}_j)|\right),
\label{Coulff}
\end{eqnarray}
where the integral of $f$ over the whole space is set to unity.
To incorporate such a finite size effect in the three-$\alpha$ system,
it is reasonable to assume the Gaussian charge form factor for
the $\alpha$ particle,
leading to the explicit form of the Coulomb potential,
\begin{align}
V_{ij}^{\rm Coul}=\frac{4e^2}{r_{ij}}{\rm erf}(\kappa r_{ij})\exp\left(-Cr_{ij}\right)
\label{Coulerf}
\end{align}
with $\kappa=0.60141$ fm$^{-1}$~\cite{Suno15}.
We will use Eq.\ (\ref{Coulerf}) unless otherwise noted.

Finally, we consider the three-$\alpha$ potential, which naturally
occurs due to the internal structure of each $\alpha$ particle. 
This potential has to be allowed for because
it is known that the empirical energies of the states close to
three-$\alpha$ threshold are not well reproduced 
only from the two-$\alpha$ potential~\cite{Suzuki02}.
As was done in Ref.~\cite{Suno15}, 
one can introduce a simplified potential
among three point $\alpha$ particles in the form of
\begin{align} 
  V_{123}^{3\alpha}=
  v_r \exp\left(-\frac{R^2}{b_r^2}\right)
  -v_a  \exp\left(-\frac{R^2}{b_a^2}\right)
\label{3apot1.eq}
\end{align}
with $R^2\equiv\sqrt{3}\sum_{i=1}^3(\bm{r}_i-\bm{x}_3)^2=\frac{\sqrt{3}}{2}x_1^2+\frac{2}{\sqrt{3}}x_2^2$.
In order to see the model dependence,
we employ two kinds of the three-$\alpha$ potential.
One is the potential that has an attractive term alone~\cite{Suno15};
the parameters are set to $v_r=0$, $v_a=152.2$ MeV, and
$b_a=2.58$ fm (Set 1) in such a way as to reproduce the empirical 
Hoyle state energy in vacuum.
We note that as given in Ref.~\cite{Suno16},
is not only the experimental charge radius of the $^{12}$C
ground state well reproduced from this Hamiltonian,
but also the calculated Hoyle state radius is consistent
with other cluster model calculations as well as the results
obtained by the Tohsaki-Horiuchi-Schuck-R\"opke
wave function~\cite{Funaki15}.
The other includes a repulsive 
term, which reasonably occurs given the Pauli principle 
among three $\alpha$ particles composed of nucleons~\cite{Saito77,Suzuki172}.
We have taken the parameters as $v_r=48.0$ MeV, $b_r=1.20$ fm,
$v_a=134$ MeV, and $b_a=2.66$ fm (Set 2), to give a totally
different version of the three-$\alpha$ potential while roughly keeping
the reproducibility of the empirical Hoyle state energy in vacuum.
Owing to difference in the structure of these two models for the 
three-$\alpha$ potential, we expect some difference in the spatial 
scale of the Hoyle state in vacuum, which in turn may lead to difference 
in the energy of the Hoyle state at nonzero $C$ in such a way that the 
larger scale, the stronger Coulomb screening.

The physical constants that we employ in this paper are 
$\hbar^2/m_\alpha$= 10.5254 MeV\,fm$^2$ and $e^2=$1.43996 MeV\,fm,
where $m_\alpha$ is the mass of an $\alpha$ particle in vacuum.

\section{Screening correction to triple alpha reactions in the 
point-charge approximation}
\label{screening.sec}

Before exhibiting the numerical solutions to the three-body problem as
described in the previous section, we follow conventional approaches to
the Coulomb screening by assuming that all the ions involved, including 
a $^{12}$C nucleus in the Hoyle state, are point charges and then estimate 
how much carbon is produced via triple alpha reactions as encountered 
in normal stars that undergo a stable burning of helium.
In such environments in which the temperature 
$T$ is of order or even higher than $10^8$ K and the mass density $\rho$
is typically $10^3$--$10^6$ g cm$^{-3}$, the Hoyle state (C$^*$)
occurs via two successive resonant reactions ($\alpha+\alpha\to{\rm Be}$ and 
${\rm Be}+\alpha\to {\rm C}^*$), where Be denotes the $^8$Be 
ground state~\cite{Clayton}. The Debye screening results 
in the Yukawa form of the Coulomb interaction among $\alpha$ particles, 
which in turn acts to enhance carbon production~\cite{Yakovlev}.
We remark that for typical conditions considered here,
the thermal kinetic energy of $\alpha$ particles is much larger
than the Coulomb energy at interparticle spacing, which is
in turn dominant over the strong force potentials
$V^{2\alpha}_{ij}$ and $V^{3\alpha}_{123}$ at the same spacing.

There are two ways of evaluating such enhancement in the carbon production.
One is a direct one in which the difference in the $Q$ value between
the screened and non-screened cases is obtained from the Coulomb
energy of the point-like Hoyle state and then is incorporated into 
the Saha prediction of the carbon production
dominated by the Boltzmann factor $e^{Q/k_B T}$~\cite{Clayton}.
Another is an indirect one in which the Coulomb correction to the 
chemical potential of each component, which is regarded as a 
point particle even for a nucleus in the Hoyle state,
is calculated in the Debye-H{\" u}ckel 
approximation and then is incorporated into 
the chemical equilibrium condition
between three $\alpha$ particles and a nucleus in the Hoyle 
state~\cite{Yakovlev,Waxman}.  
As far as the system is sufficiently hot to become
a weakly coupled, non-degenerate plasma that is charge neutral,
both approaches have to give a consistent result for the enhancement
in the carbon production.
We remark that corrections due to the electron
Fermi degeneracy and/or the strong Coulomb coupling would make 
the $\alpha$-$\alpha$ Coulomb interaction deviate from the simple 
Debye-screened form~\cite{Ichimaru}.

As for the first approach, all we have to do is to give the $C$ 
value appropriately. In the case of the Debye screening, 
$C^{-1}$ is the Debye screening length defined as
\begin{equation}
\lambda_D=\left[\frac{k_B T}{4\pi e^2(n_e+\sum_i n_i Z_i^2)}\right]^{1/2},
\label{Debye1}
\end{equation}
where $n_i$ and $Z_i$ are the averaged number density and charge number of 
ions of species $i$, and $n_e$ is the averaged number density of electrons.
Due to charge neutrality, $n_e=\sum_i Z_i n_i$ is satisfied.
We can estimate the value of $\lambda_D$ by assuming
that hydrogen is exhausted and that $\sum_i n_i Z_i^2$ is dominated by 
$\alpha$ particles ($i=\alpha$).  The latter assumption is validated if 
one notes the fact that under chemical equilibrium, $n_{{\rm C}^*}$ is 
proportional to $e^{Q/k_B T}$, which is generally negligible. 
In the range of $T$ and $\rho$ as considered here, $\lambda_D$ is 
of order $10^3$--$10^4$ fm.  

Eventually, this $\lambda_D$ determines corrections to the Coulomb potential
of a quantum system of three $\alpha$ particles as 
\begin{equation}
\Delta V_C=\sum_{j<k} \frac{4e^2}{r_{jk}}e^{-r_{jk}/\lambda_D}
          -\sum_{j<k} \frac{4e^2}{r_{jk}},
\label{DEC}
\end{equation}
where $r_{jk}$ is the distance operator between the $j$th and $k$th $\alpha$
particles.  The expectation value of $\Delta V_C$, which 
can be obtained in the present 
three-body calculations as $\Delta E_C=\langle \Delta V_C \rangle$, 
gives rise to decrease in the mass
of the Hoyle state and hence increase in the $Q$ value.  
Since the distance between the fusing particles is generally 
far shorter than $\lambda_D$, we obtain, by using the 
Taylor expansion with respect to $\langle r_{jk} \rangle$,
\begin{equation}
\Delta E_C=-\frac{12e^2}{\lambda_D}+O(\langle r_{jk}\rangle).
\label{DEC1}
\end{equation}
Equation (\ref{DEC1}) suggests that both in the weak 
screening limit and in the zero-size limit of the Hoyle state,
the increase in the $Q$ value amounts to $\frac{12e^2}{\lambda_D}$.
We remark in passing that the number of electrons remain unchanged by 
the triple alpha reaction
and that the gamma decay of the Hoyle state (two-photon processes) 
is not considered here because the lower-lying carbon states are not
always described in terms of three $\alpha$ particles.

As for the second approach, we first write down the Helmholtz free energy
density of the system as 
\begin{equation}
f=f_0+f_{\rm DH},
\label{ftot}
\end{equation}
where
\begin{eqnarray}
f_0 &=& n_e m_e c^2 
-n_e k_B T\left\{\ln\left[\frac{2}{n_e}
   \left(\frac{m_e k_B T}{2\pi\hbar^2}\right)^{3/2}\right]+1\right\}
\nonumber \\ & &
+\sum_i n_i m_i c^2
-\sum_i n_i k_B T\left\{\ln\left[\frac{g_i}{n_i}
   \left(\frac{m_i k_B T}{2\pi\hbar^2}\right)^{3/2}\right]+1\right\}
\label{f0}
\end{eqnarray}
with the electron ($i$ ion) rest mass $m_e$ $(m_i)$
and the number of internal degrees of freedom of $i$ ions $g_i$, 
is the ideal gas part of the free energy density, and
\begin{equation}
f_{\rm DH} = -(n_e+\sum_i n_i Z_i^2)\frac{e^2}{3\lambda_D}
\label{fdh}
\end{equation}
is the lowest order Coulomb correction to $f_0$, i.e., the 
Debye-H{\" u}ckel term appropriate for a multi-component classical plasma.
From this free energy density, one can derive the chemical potential of 
electrons and of $i$ ions as
\begin{equation}
\mu_e=m_e c^2 - k_B T \ln\left[\frac{2}{n_e}
   \left(\frac{m_e k_B T}{2\pi\hbar^2}\right)^{3/2}\right]
   -\frac{e^2}{2\lambda_D}
\label{mue}
\end{equation}
and 
\begin{equation}
\mu_i=m_i c^2 - k_B T \ln\left[\frac{g_i}{n_i}
   \left(\frac{m_i k_B T}{2\pi\hbar^2}\right)^{3/2}\right]
   -\frac{Z_i^2 e^2}{2\lambda_D},
\label{mui}
\end{equation}
respectively.  The last term of the right side in Eq.\ (\ref{mui}) 
corresponds to the Coulomb correction, $\mu^{\rm Coul}_i$, to the chemical 
potential of $i$ ions.

We then apply the chemical potentials given by Eqs.\ (\ref{mue}) 
and (\ref{mui}) to the chemical equilibrium condition for
$3\alpha \leftrightarrow {\rm C}^*$,
\begin{equation}
3\mu_\alpha = \mu_{{\rm C}^*}.
\label{chemeq}
\end{equation}
Note that $\mu_e$ does not come in because the triple
alpha process involves no beta process.
We thus obtain
\begin{equation}
n_{{\rm C}^*}=n_\alpha^3 e^{Q_0/k_B T} 
\lambda_\alpha^9 \lambda_{{\rm C}^*}^{-3}
e^{12e^2/\lambda_D k_B T},
\label{Cpro}
\end{equation}
where $Q_0=(3m_\alpha-m_{{\rm C}^*})c^2$ is the $Q$ value in the
ideal gas limit, 
$\lambda_i=\sqrt{2\pi}\hbar/\sqrt{m_i k_B T}$ is the thermal de Broglie 
wavelength, and $g_{{\rm C}^*}=g_\alpha=1$.  In the absence of screening,
Eq.\ (\ref{Cpro}) reduces to the Saha prediction of the carbon production.
The Debye screening induces the factor $e^{12e^2/\lambda_D k_B T}$ via
$3\mu^{\rm Coul}_\alpha-\mu^{\rm Coul}_{{\rm C}^*}=12 e^2 / \lambda_D$,
which is consistent with the first approach that predicts increase 
in the $Q$ value by $12 e^2 / \lambda_D$ in the weak screening limit.

\section{Results and discussions}
\label{results.sec}

 We now proceed to exhibit the numerical results for 
the energy and size of the Hoyle state in weakly-coupled, 
thermal plasmas.   The former will then be compared with the point-charge 
prediction of the $Q$ value shift as given in the previous section.

\subsection{Coulomb screening effects on the three-$\alpha$ system}

\begin{figure}[ht]
\begin{center}
\epsfig{file=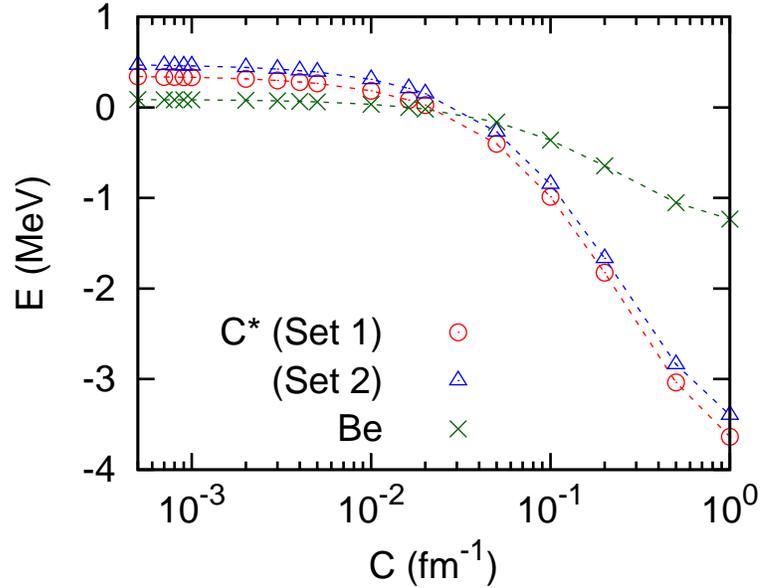,scale=2}
\caption{Energy of the screened Hoyle
state of the three-$\alpha$ system 
calculated with respect to the three-$\alpha$ threshold
as a function of the screening factor $C$. 
The result for the screened ground state of 
$^{8}$Be is also plotted for comparison.}
\label{energy.fig}
\end{center}
\end{figure}

In the present framework based on the Hamiltonian 
(\ref{hamiltonian}), the lowest energy state $(n=0)$ corresponds to 
the ground state of $^{12}$C, while the Hoyle state appears 
as a resonant/bound excited state.  Since the decay width of 
the Hoyle state is small, such a resonant state
can be essentially described
as a bound state~\cite{Horiuchi08, Horiuchi13a, Horiuchi13b}.
In fact, in the two-$\alpha$ system,
the ground state energy of $^8$Be is obtained as 88.8 keV.
With the Set 1 Hamiltonian $(C=0)$,
the energies of the $^{12}$C ground state and the Hoyle state
are $-9.40$ and $0.349$ MeV, respectively,
which are consistent with the results of Ref.~\cite{Suno15}, while
the Set 2 Hamiltonian ($C=0$) gives $-$8.99 and 0.475 MeV
for the ground and Hoyle states, respectively.

Figure~\ref{energy.fig} plots
the energies of the screened Hoyle state
and the ground state of $^{8}$Be with respect to the three-$\alpha$ 
threshold, $E_{{\rm C}^*}$ and $E_{\rm Be}$, evaluated 
as a function of the Coulomb screening factor $C$.
As expected, the Hoyle state energy decreases 
with increasing the Coulomb screening factor $C$ 
and eventually approaches the three-$\alpha$ energy 
calculated in the absence of the Coulomb term
in the Hamiltonian. When $C=0.0162$ fm$^{-1}$, 
$E_{\rm Be}$ becomes $\sim 10^{-5}$ MeV,
which suggest that the condition for appearance
of the Efimov state is met. We nevertheless find
that for both Hamiltonians
the Hoyle state is still in a resonance state with $E_{{\rm C}^*}=0.082$ (0.210) MeV
for Set 1 (Set 2), excluding
that the hypothesis that the Hoyle state is bound by the Efimov attraction.
Beyond $C\sim 0.05$ fm$^{-1}$, 
we observe that the Hoyle state appears as a bound state, that is,
the energy becomes below the $^8$Be one, and that the 
asymptotic energy ($C=\infty$) calculated with Set 2 is $-3.62$ MeV,
being slightly higher than that with Set 1 ($-3.89$ MeV)
because Set 2 includes the repulsive component
in the three-$\alpha$ potential.
In this situation, however, the screening is too strong for the 
Debye-H{\" u}ckel approximation to be valid.
We can also observe the behavior of the energy shift 
at small $C$ does not depend strongly on the choice of 
the three-$\alpha$ potential, which will be discussed quantitatively
later in this section.  

\begin{figure}[ht]
\begin{center}
\epsfig{file=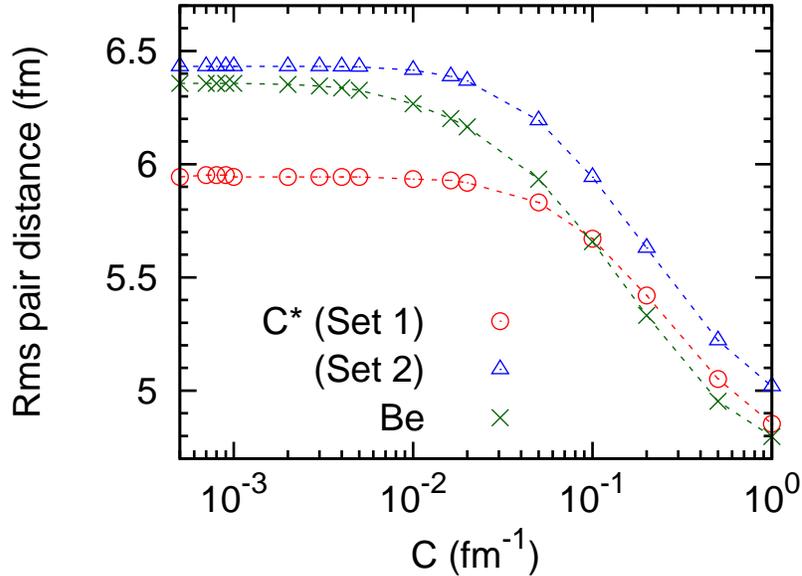,scale=2}
\caption{Root-mean-square (rms) pair distances 
of the screened Hoyle state  and of the screened $^8$Be 
ground state calculated as a function of the screening 
factor $C$.}
\label{rms.fig}
\end{center}
\end{figure}

To examine more details of the correlated motion of the three-$\alpha$ 
system, we calculate the root-mean-square (rms) pair distance defined by 
$d^{(n)}=\sqrt{\left<\Psi^{(n)}\right|\bm{x}_1^2\left|\Psi^{(n)}\right>.}$
Figure~\ref{rms.fig} plots the results for the rms pair distance
of the Hoyle state as a function of $C$, together with
those of the ground state of $^8$Be for comparison.
We remark that a relatively large
  rms pair distance 6.3 fm of $^{8}$Be
  is obtained as compared to
  the {\it ab initio} calculation 4.8 fm~\cite{Carlson15},
although the wave function calculated here well reproduces
the empirical energy and decay width of $^{8}$Be~\cite{Suno15}.
For Set 1, as long as $C$ is small, the rms pair distance of the Hoyle 
state is significantly shorter than that of the $^8$Be 
ground state due to stronger binding in the three-$\alpha$ system.
At a critical $C$ where the Hoyle state becomes bound,
the rms pair distance of the $^{8}$Be ground state becomes 
so short as to coincide with that of the Hoyle state.
This is not the case with Set 2, which provides the Hoyle state 
with a pair distance that is longer than not only the same quantity
calculated from Set 1, but also the $^8$Be result for any 
positive $C$.  This behavior comes from 
the repulsive component of the three-$\alpha$ potential in Set 2.
In fact, the rms pair distance of the $^{8}$Be ground state 
with $C=\infty$ becomes 4.75 fm, while
that of C$^*$ is 4.75 fm for Set 1 and 4.92 fm for Set 2.
Such repulsion also acts to enhance the rms radius of 
the Hoyle state, which can be measured from the center of mass
of the system as 3.43 (2.74) fm for Set 1 and 3.71 (2.84) fm for Set 2, 
respectively, in the case of $C=0$ ($C=\infty$).

\subsection{Screening-induced enhancement of carbon production}

Let us now consider a realistic situation in which the value 
of $C$ is set to the inverse of the Debye screening length 
(\ref{Debye1}), $\lambda_D^{-1}$.  The use of $\lambda_D$ is 
strictly applicable to a plasma in which all the components 
(ions and electrons) behave like a nearly ideal, thermal gas.
For example, at the highest density of interest here, the 
Fermi degeneracy can play a role in modifying the 
present description of the screening correction to the 
Coulomb interaction as suggested in Sec.\ \ref{screening.sec}.

Next, we reestimate screening-induced enhancement of 
the carbon production in normal stars by allowing for the
spatial structure of the Hoyle state as precisely
evaluated in the previous subsection.  To do so, we follow
the same line of argument of the direct approach 
shown in Sec.\ \ref{screening.sec}.  Instead of taking 
the zero-size limit as in Sec.\ \ref{screening.sec}, 
we just substitute $E_{{\rm C}^*}(C)$ into $Q$ and 
thereby estimate the screening-induced $Q$ value shift 
and enhancement factor as 
$\Delta Q(C)=E_{{\rm C}^*}(C=0)-E_{{\rm C}^*}(C)$
and $e^{\Delta Q(C)/k_B T}$, respectively.

\begin{figure}[ht]
\begin{center}
\epsfig{file=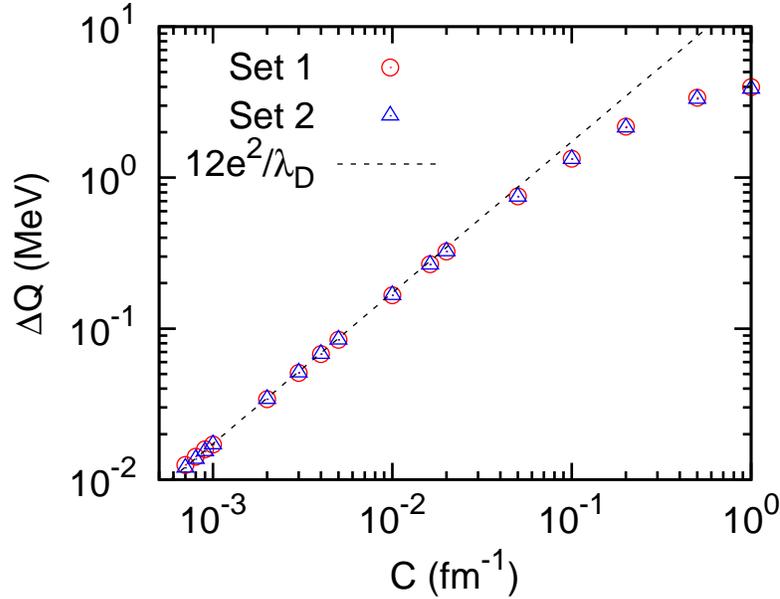, scale=2}  
\caption{Screening-induced $Q$ value shifts of the Hoyle state 
calculated as a function of $C$ via 
$\Delta Q=E_{{\rm C}^*}(C=0)-E_{{\rm C}^*}(C)$.
The dashed line denotes the logarithm of the screening-induced
enhancement factor with $\lambda_D$ being identified with $C^{-1}$.}
\label{fig:screen}
\end{center}
\end{figure}

Finally, we compare the resultant $Q$ value shift
due to the screening, $\Delta Q(C)$,
with the conventional prediction, $12e^2/\lambda_D$, obtained 
for point charges.  In Fig.\ \ref{fig:screen} we show
such a comparison by regarding $C$ as $\lambda_D^{-1}$.
The obtained $\Delta Q$ values are insensitive to the 
three-$\alpha$ potential and hence the size of the Hoyle state.
Both for Set 1 and Set 2, the results agree with the
conventional prediction in the limit of $C\to 0$, as they should.  
For typical $\lambda_D$, such an agreement seems to be intact.
We also evaluate $\Delta Q(C)$ with the three-$\alpha$
  calculations using the point-charge
  Coulomb potential of Eq.~(\ref{Coul}) and find that
  the results are virtually the same
  as the ones presented in Fig.~\ref{fig:screen}.

Despite such a good agreement, there has to be a difference
in the screening-induced $Q$ value shift between the finite-size 
and zero-size cases of the Hoyle state. 
This difference,
as discussed in Sec.\ \ref{model.sec}, is expected to depend on
the model adopted for the three-$\alpha$ potential because the two
models give an appreciable difference in the prediction of the spatial 
scale of the Hoyle state as shown in Fig.\ \ref{rms.fig}.
To estimate the difference in the $Q$ value shift, however, it
may be inappropriate to use the present results
  of $\Delta Q(C)$ with the Yukawa form of the Coulomb potential
  because even in the weak coupling and classical limit, the Debye-H\"uckel 
approximation holds only for a description of the long-range Coulomb 
screening. In fact, within this approximation, the radial distribution 
function, $g(r)$, for $\alpha$ particles, i.e., the probability of finding 
another $\alpha$ particle at a distance of $r$ from the origin at
which an $\alpha$ particle is already located, is known to be 
negative near $r=0$ and hence unphysical \cite{Ichimaru2}.  
In the case of the triple alpha reactions, 
the fusing $\alpha$ particles are inevitably located in the 
immediate vicinity of the partner.  It is thus necessary to 
properly take into account the short-range spatial correlation.
Here we estimate such an effect on the present $Q$-value shift.
To do so, it is convenient to note that for thermal plasmas one can 
generally express $g(r)e^{4e^2/rk_B T}$ in a power series of $r^2$ 
near $r=0$ \cite{Widom}.  Then it is reasonable to define the effective 
potential $w(r)$ between two $\alpha$ particles via
$g(r)=e^{-w(r)/k_B T}$.  According to Ref.\ \cite{Jancovici}, $w(r)$ can
be expanded as
\begin{eqnarray}
w(r)&=&\frac{4e^2}{r}-\frac{4e^2}{\lambda_D}+\frac{1}{4} \frac{4e^2}{a_\alpha}\left[\frac{r}{a_\alpha}\right]^2+O(r^4)
\label{effpot}
\end{eqnarray}
with $a_\alpha=(3/4\pi n_\alpha)^{1/3}$.
The second term of the right side
corresponds to $-(2\mu^{\rm Coul}_\alpha-\mu^{\rm Coul}_{{\rm Be}})$, i.e., 
the conventional point-charge prediction of the 
screening-induced $Q$ value shift for $\alpha+\alpha\to{\rm Be}$ 
under chemical equilibrium $2\alpha \leftrightarrow {\rm Be}$, 
while the third term comes from two closely separated $\alpha$ 
particles in the uniform electron background.  For typical 
separation as depicted in Fig.\ \ref{rms.fig} as well as 
for typical $T$ and $\rho$, the ratio of the third to second 
term is only of order $10^{-7}$.  The usage of $w(r)$ instead of
the Yukawa potential in Eq.\ (\ref{DEC}) would thus reproduce 
the screening-induced $Q$ value shift of the triple alpha
process in the point-charge approximation, $12e^2/\lambda_D$, 
while adding an $O(10^{-7})$ correction due to the finite 
size effect of the Hoyle state.  This implies that the difference 
in such a $Q$ value shift between the two models for the 
three-$\alpha$ potential would also be negligible.  
We remark in passing that in the present estimate of the terms 
beyond the second one in Eq.\ (\ref{effpot}), possible corrections 
due to the electron screening, the strong force potential 
$V_{ij}^{2\alpha}$, and the quantum nature of fusing $\alpha$ 
particles are ignored.

\section{Conclusion}
\label{conclusion.sec}

In this paper, we have revisited the Coulomb screening 
correction to the $Q$ value of the triple alpha process
in weakly-coupled, thermal plasmas by newly obtaining 
the precise three-$\alpha$ wave function 
within the Debye-H\"uckel approximation.  Through
variational calculations that incorporate
a finite size effect of the Hoyle state, we find 
that the conventional point-charge analysis gives a very 
good estimate of the screening-induced $Q$ value shift in 
normal stars that undergo a stable burning of helium.
We also find that our three-$\alpha$ calculation
within the Yukawa form of the Coulomb potential
does not support the conjecture made in Refs.~\cite{Jensen95, Higa08}
that the Hoyle state could emerge from the Efimov state.

Many questions nevertheless remain.  It would be 
straightforward to perform the same kind of three-body 
calculations by considering a more realistic situation, 
e.g., by using the effective Coulomb potential (\ref{effpot}) 
instead of the Yukawa potential (\ref{Coul}).
Estimates of the $Q$ value shift in different environments 
as may be encountered in X-ray bursting, accreting neutron 
stars would also be interesting.

\section*{Acknowledgment}

This work was in part supported by JSPS KAKENHI Grants
Nos.\ 18K03635, 18H01211, 18H04569, 18H05406, and 19H05140.
We acknowledge the collaborative research program 2020, 
information initiative center, Hokkaido University.
MTY thanks the Brazilian agencies Funda\c{c}\~ao de Amparo 
\`a Pesquisa do Estado de S\~ao Paulo-FAPESP Grants 
No.\ 2019/00153-8 and Conselho Nacional de
Desenvolvimento Cient\'ifico e Tecnol\'ogico-CNPq Grant 
No.\ 303579/2019-6.


\begin{thebibliography}{99}
\bibitem{Beck} C. Beck (Ed.), Clusters in Nuclei, vol 1 2010, vol 2 2012,
vol 3 2014, Springer Berlin, and references therein.
\bibitem{Ishikawa13} S. Ishikawa, Phys. Rev. C {\bf 87}, 055804 (2013).
\bibitem{Suno15} H. Suno, Y. Suzuki, and P. Descouvemont, 
  Phys. Rev. C {\bf 91}, 014004 (2015).
  \bibitem{Akahori15} T. Akahori, Y. Funaki, and K. Yabana,
  Phys. Rev. C {\bf 92}, 022801(R) (2015).
\bibitem{Suno16} H. Suno, Y. Suzuki, and P. Descouvemont, Phys. Rev. C {\bf 94}, 054607 (2016).  
\bibitem{Salpeter52} E.~E. Salpeter, Astrophys. J., 115, 326 (1952).
\bibitem{Hoyle54} F. Hoyle, Astrophys. J. Suppl. Ser., {\bf 1}, 12 (1954).
\bibitem{Dunbar53} D.~N.~F. Dunbar, R.~E. Pixley, W.~A. Wenzel, Phys. Rev., {\bf 92}, 649 (1953).
\bibitem{Lewin93} W.~H.~G. Lewin, J. van Paradijs, and R.~E. Taam,
Space Sci.\ Rev., {\bf 62}, 223 (1993).
\bibitem{Salpeter54} E.-E. Salpeter, Aust. J. Phys., {\bf 7}, 373 (1954).
\bibitem{Varga95} K. Varga and Y. Suzuki, Phys. Rev. C {\bf 52}, 2885 (1995).
\bibitem{SVM} Y. Suzuki and K. Varga,  {\it Stochastic Variational Approach to Quantum-Mechanical
Few-Body Problems}, Lecture Notes in Physics, Vol. m54 (Springer, Berlin, 1998). 
\bibitem{Jensen95} A.~S. Jensen, D.~V. Fedorov, K. Langanke, and H.~M. Muller,
Proceedings of the International Conference on Exotic Nuclei and Atomic Masses (ENAM5), Ed. by M. de Saint Simon and O. Sorlin (Ed. Frontieres, Gif-sur-Yvette, 1995), p. 677.
\bibitem{Higa08} R. Higa, H.-W. Hammer, and U. van Kolck, 
Nucl. Phys. {\bf A 809}, 171 (2008).
\bibitem{Naidon17} P. Naidon and S. Endo, Rep.\ Prog.\ Phys.\ {\bf 80}, 056001
  (2017).
\bibitem{Mitroy13} J. Mitroy, S. Bubin, W. Horiuchi, Y. Suzuki, 
L. Adamowicz, W. Cencek, K. Szalewicz, J. Komasa, D. Blume, and K. Varga,
Rev. Mod. Phys. {\bf 85}, 693-749 (2013).
\bibitem{Suzuki17} Y. Suzuki and W. Horiuchi,
{\it Emergent Phenomena in Atomic Nuclei from Large-scale Modeling: A Symmetry-Guided Perspective''} (World Scientific, Singapore, 2017),  Chap. 7, pp. 199-227.
\bibitem{Suzuki08} Y. Suzuki, W. Horiuchi, M. Orabi, and K. Arai,
Few-Body Syst. {\bf 42}, 33 (2008).

\bibitem{AB} S. Ali and A.~R. Bodmer, Nucl. Phys. {\bf 80}, 99 (1966). 
\bibitem{BFW} B. Buck, H. Friedrich, and C. Wheatley, 
Nucl. Phys. {\bf A 275}, 246 (1977).
\bibitem{Theeten06} M. Theeten, D. Baye, and P. Descouvemont,
Phys. Rev.  C {\bf 74}, 044304 (2006).
\bibitem{Fedorov96} D.~V. Fedorov and A.~S. Jensen, Phys. Lett. {\bf B 389},
  631 (1996).
\bibitem{Ajzenberg90} F. Ajzenberg-Selove, Nucl. Phys. {\bf A 506}, 1 (1990).  
\bibitem{Suzuki02} Y. Suzuki and M. Takahashi,
  Phys. Rev. C {\bf 65}, 064318 (2002).
  \bibitem{Funaki15} Y. Funaki, H. Horiuchi, and A. Tohsaki, Prog. Part. Nucl. Phys. {\bf 82},
  78 (2015).
\bibitem{Saito77} S. Saito, Prog. Theor. Phys. Suppl. {\bf 62}, 11 (1977).
\bibitem{Suzuki172} Y. Suzuki and W. Horiuchi, Phys. Rev. C {\bf 95}, 044320 (2017).
\bibitem{Carlson15} J. Carlson, S. Gandolfi, F. Pederiva, Steven C. Pieper, R. Schiavilla, K. E. Schmidt,
  and R. B. Wiringa, Rev. Mod. Phys. {\bf 87}, 1067 (2015).
\bibitem{Clayton} D.~D. Clayton, 
{\it Principles of Stellar Evolution and Nucleosynthesis} (McGraw-Hill, New York, 1968).
\bibitem{Yakovlev} D.G. Yakovlev and D.A. Shalybkov, Astrophys.\ Space
Phys.\ Rev.\ {\bf 7}, 311 (1989).
\bibitem{Waxman} D. Kushnir, E. Waxman, and A.~I. Chugunov, Mon.\ Not.\ R. 
Astron.\ Soc.\ {\bf 486}, 449 (2019).
\bibitem{Ichimaru} S. Ichimaru, Rev.\ Mod.\ Phys.\ {\bf 65}, 255 (1993).
\bibitem{Horiuchi08} W. Horiuchi and Y. Suzuki, Phys. Rev. C {\bf 78}, 034305 (2008).
\bibitem{Horiuchi13a} W. Horiuchi and Y. Suzuki, Few-Body Syst. {\bf 54}, 2407 (2013).
\bibitem{Horiuchi13b} W. Horiuchi and Y. Suzuki, Phys. Rev. C {\bf 87}, 034001 (2013).
\bibitem{Ichimaru2} S. Ichimaru, Rev.\ Mod.\ Phys.\ {\bf 54}, 1017 (1982).
\bibitem{Widom} B. Widom, J. Chem.\ Phys.\ {\bf 39}, 2808 (1963).
\bibitem{Jancovici} B. Jancovici, J. Stat.\ Phys.\ {\bf 17}, 357 (1977).
\end{thebibliography}
\end{document}